\documentclass{PoS}

\usepackage{amsmath,amssymb}
\usepackage{graphicx}
\usepackage{bm}
\usepackage{pgf,tikz}
\usetikzlibrary{arrows}
\usetikzlibrary[patterns]
\definecolor{zzttqq}{rgb}{0.6,0.2,0}
\definecolor{cqcqcq}{rgb}{0.75,0.75,0.75}
\usepackage{float}

\title{
\vspace*{-1.5cm}
\begin{flushright}
\texttt{\footnotesize CERN-PH-TH-2015-309}\\
\end{flushright}
\vspace*{0.5cm}
Aspects of topological actions on the lattice}

\ShortTitle{Topological actions}

\author{Oscar {\AA}kerlund \\
  Institut f\"ur Theoretische Physik, ETH Zurich, CH-8093 Z\"urich, Switzerland\\
        E-mail: \email{oscara@phys.ethz.ch}}

\author{\speaker{Philippe de Forcrand}\\
  Institut f\"ur Theoretische Physik, ETH Zurich, CH-8093 Z\"urich, Switzerland\\
  CERN, Physics Department, TH Unit, CH-1211 Geneva 23, Switzerland\\
        E-mail: \email{forcrand@phys.ethz.ch}}

\abstract{
We consider a lattice action which forbids large fields, and which remains invariant under smooth
deformations of the field. Such a ``topological'' action depends on one parameter, the field cutoff,
but does not have a classical continuum limit as this cutoff approaches zero. 
We study the properties of such an action in $4d$ compact $U(1)$ lattice gauge theory,
and compare them with those of the Wilson action. In both cases, we find a weakly 
first-order transition separating a confining phase where monopoles condense, and
a Coulomb phase where monopoles are exponentially suppressed. We also find
a different, critical value of the field cutoff where monopoles completely disappear.
Finally, we show that a topological action simplifies the measurement of the free energy.
}

\FullConference{The 33rd International Symposium on Lattice Field Theory\\
		14 -18 July 2015\\
		Kobe International Conference Center, Kobe, Japan}

\begin{document}

\section{Introduction}

At least two strategies can be adopted to improve the approach of a lattice
action to the continuum limit. The most widely considered is the Symanzik
strategy~\cite{Symanzik}, where irrelevant operators of increasing dimension are added
to the standard action, with coefficients determined perturbatively or not.
Another, less common, strategy is motivated by the observation that the
approach to the continuum behavior is faster for weaker fields: suppressing
strong fields will improve the action in a non-parametric way. Indeed,
the Manton action~\cite{Manton}, or renormalization-group improved actions like the
``perfect'' action~\cite{perfect} all suppress large fields.

A radical implementation of this second strategy consists of imposing
a cutoff on the amount of local excitation, the earliest example being
the positive-plaquette action~\cite{positive}. Recently, a combination of Wilson-type
action and cutoff constraint has been shown to display much improved scaling
in a spin model~\cite{Bern_XY}. Very recently, a similar approach has been
applied to a gauge theory~\cite{Bern_SU2}. 

This leaves open the question of the approach to the continuum limit of
a constraint-only action, namely an action $S$ such that $\exp(-S) = 1$
if every local constraint is satisfied, $0$ otherwise. Such an action,
which is invariant under small deformations of the field, is for this
reason called ``topological''. While it is intuitively clear that the
correlation length will keep increasing as the constraint selects 
weaker and weaker fields, it is not at all obvious which continuum theory
will be approached: a gradient expansion of the lattice action is not
possible, since the action is either zero or infinite.
Nevertheless, it has been shown numerically~\cite{top_XY} that in a 
$2d$ $XY$ spin model,
the same Coulomb phase is obtained as with the standard  action.
Here, we study the case of a $4d$ $U(1)$ gauge theory, and compare
the Wilson action $S_W = - \beta \sum_P \cos \theta_P$ with the 
topological action 
\begin{equation}
\exp(-S_{\rm top}) = \prod_P \Theta(\delta - |\theta_P|)
\end{equation}
where $\theta_P$ is the angle of plaquette $P$ and $\delta \in [0,\pi]$
is the cutoff on $\theta_P$. Most of the results presented here have
appeared in \cite{top_U1}.

\section{Monopoles and ergodicity}

It is well-known that, using the Wilson action, the $4d$ $U(1)$ lattice
gauge theory has two phases: Coulomb at weak coupling, confining at
strong coupling, with a weak first-order transition at $\beta_c \approx 1.01\cdots$.
Moreover, in the confining phase magnetic monopoles condense~\cite{DeG_T}.
With a topological action, the existence of monopoles is governed by the
cutoff angle $\delta$. As shown Fig.~1 (left), a total flux of $2\pi$ 
through the 6 faces of a cube, which characterizes a monopole, requires $\delta$ to be at least $\pi/3$.
Thus the monopole density vanishes as $\delta\to \frac{\pi}{3}^+$, and
is strictly zero for $\delta < \frac{\pi}{3}$. This non-analyticity of
the monopole density causes a phase transition at $\delta=\pi/3$, which
one may guess coincides with the Coulomb-confining transition. This guess
is wrong as we will show below.

The numerical study of the regime $\delta < \pi$ requires care.
As shown Fig.~1 (middle), the usual single-link update will not succeed
in changing the monopole flux by $2\pi$, and thus is not ergodic.
To maintain ergodicity requires updating at least two links simultaneously,
as shown Fig.~1 (right).

\begin{figure}
\centerline{
\includegraphics[width=2.3cm,viewport= 0 20 200 220]{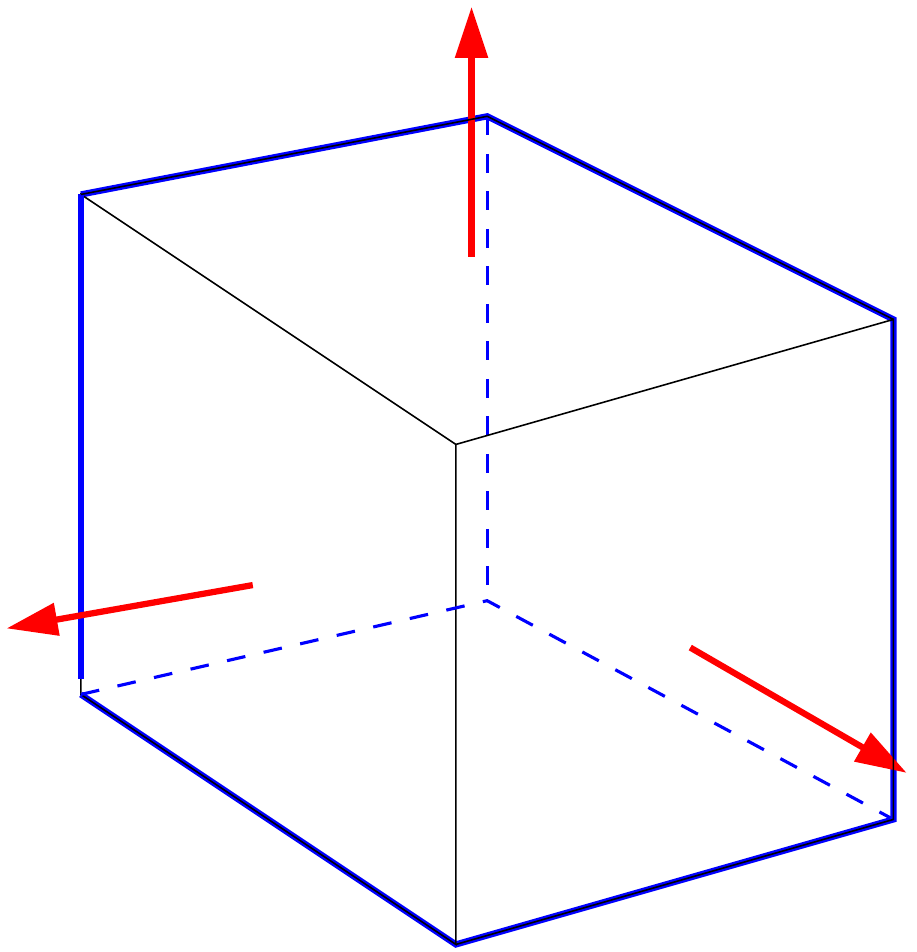}
\hspace*{-0.5cm}
\begin{tikzpicture}[line cap=round,line join=round,>=triangle 45,x=0.4cm,y=0.4cm]
\clip(-4,-2.5) rectangle (16,4.0);
\draw (2,2)-- (2,-2);
\draw (2,-2)-- (6,-2);
\draw (2,2)-- (6,2);
\draw (6,-2)-- (6,2);
\draw (2,2)-- (4,4);
\draw (4,4)-- (8,4);
\draw (6,2)-- (8,4);
\draw (8,4)-- (8,0);
\draw (8,0)-- (6,0);
\draw [dash pattern=on 3pt off 3pt] (6,0)-- (4,0);
\draw [dash pattern=on 3pt off 3pt] (4,0)-- (2,-2);
\draw [dash pattern=on 3pt off 3pt] (4,0)-- (4,4);
\draw (8,-2)-- (12,-2);
\draw (8,-2)-- (10,0);
\draw (10,0)-- (12,0);
\draw [dash pattern=on 3pt off 3pt] (12,0)-- (14,0);
\draw [line width=2.8pt] (12,-2)-- (14,0);
\draw (12,-2)-- (12,2);
\draw (12,2)-- (14,4);
\draw (14,4)-- (14,0);
\draw (4.54,1.34) node[anchor=north west] {$\pi$};
\draw (10.57,1.64) node[anchor=north west] {$\pm\pi$};
\draw (1,-10)-- (1,-6);
\draw (5,-6)-- (1,-6);
\draw (1,-6)-- (3,-4);
\draw (3,-4)-- (7,-4);
\draw (7,-4)-- (5,-6);
\draw (1,-10)-- (3,-8);
\draw (3,-8)-- (7,-8);
\draw (7,-4)-- (7,-8);
\draw (3,-8)-- (3,-6);
\draw [dash pattern=on 3pt off 3pt] (3,-4)-- (3,-6);
\draw (9,-10)-- (13,-10);
\draw [line width=2.8pt] (13,-6)-- (13,-10);
\draw (9,-6)-- (9,-10);
\draw (9,-6)-- (13,-6);
\draw (13,-6)-- (15,-4);
\draw (15,-4)-- (15,-8);
\draw [line width=2.8pt] (15,-8)-- (13,-10);
\draw [dash pattern=on 3pt off 3pt] (9,-10)-- (11,-8);
\draw [dash pattern=on 3pt off 3pt] (11,-8)-- (15,-8);
\draw (4.23,-6.39) node[anchor=north west] {$\pi$};
\draw (11.27,-4.56) node[anchor=north west] {$\pm\pi$};
\end{tikzpicture}
\hspace*{-1.8cm}
\begin{tikzpicture}[line cap=round,line join=round,>=triangle 45,x=0.4cm,y=0.4cm]
\clip(-4,-10.5) rectangle (16,-4.0);
\draw (2,2)-- (2,-2);
\draw (2,-2)-- (6,-2);
\draw (2,2)-- (6,2);
\draw (6,-2)-- (6,2);
\draw (2,2)-- (4,4);
\draw (4,4)-- (8,4);
\draw (6,2)-- (8,4);
\draw (8,4)-- (8,0);
\draw (8,0)-- (6,0);
\draw [dash pattern=on 3pt off 3pt] (6,0)-- (4,0);
\draw [dash pattern=on 3pt off 3pt] (4,0)-- (2,-2);
\draw [dash pattern=on 3pt off 3pt] (4,0)-- (4,4);
\draw (8,-2)-- (12,-2);
\draw (8,-2)-- (10,0);
\draw (10,0)-- (12,0);
\draw [dash pattern=on 3pt off 3pt] (12,0)-- (14,0);
\draw [line width=2.8pt] (12,-2)-- (14,0);
\draw (12,-2)-- (12,2);
\draw (12,2)-- (14,4);
\draw (14,4)-- (14,0);
\draw (4.54,1.34) node[anchor=north west] {$\pi$};
\draw (10.57,1.64) node[anchor=north west] {$\pm\pi$};
\draw (1,-10)-- (1,-6);
\draw (5,-6)-- (1,-6);
\draw (1,-6)-- (3,-4);
\draw (3,-4)-- (7,-4);
\draw (7,-4)-- (5,-6);
\draw (1,-10)-- (3,-8);
\draw (3,-8)-- (7,-8);
\draw (7,-4)-- (7,-8);
\draw (3,-8)-- (3,-6);
\draw [dash pattern=on 3pt off 3pt] (3,-4)-- (3,-6);
\draw (9,-10)-- (13,-10);
\draw [line width=2.8pt] (13,-6)-- (13,-10);
\draw (9,-6)-- (9,-10);
\draw (9,-6)-- (13,-6);
\draw (13,-6)-- (15,-4);
\draw (15,-4)-- (15,-8);
\draw [line width=2.8pt] (15,-8)-- (13,-10);
\draw [dash pattern=on 3pt off 3pt] (9,-10)-- (11,-8);
\draw [dash pattern=on 3pt off 3pt] (11,-8)-- (15,-8);
\draw (4.23,-6.39) node[anchor=north west] {$\pi$};
\draw (11.27,-4.56) node[anchor=north west] {$\pm\pi$};
\end{tikzpicture}
}
\caption{{\em Left}: A magnetic monopole is identified via its $2\pi$ 
magnetic flux exiting an elementary cube. Thus, it requires some plaquette
angles to reach or exceed $\pi/3$. If the action forbids such plaquette
angles, monopoles are absent. {\em Middle}: Creating or destroying a 
monopole requires changing the magnetic flux exiting an elementary cube 
by $2\pi$.
A $2\pi$ flux change via a single-link update implies changing by $\pi$
the angles of two plaquettes. Thus, if the action forbids plaquette angles
larger than $\pi$ and a single-link update is used, monopoles will be frozen 
and ergodicity will be lost.  {\em Right}: This problem
disappears, and ergodicity is restored, if {\em two} links are updated
simultaneously.
}
\end{figure}
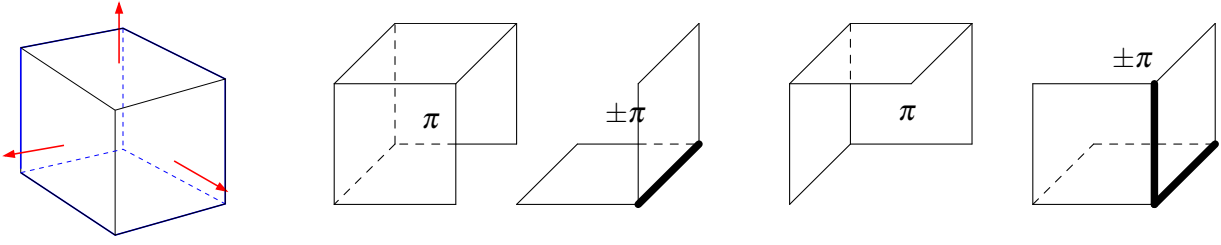

\newpage

\section{Helicity modulus and confining/Coulomb phase transition}

\noindent
The helicity modulus $h$
measures the response of the system to a twist in the boundary conditions:
\begin{equation}
h \equiv \left.\frac{\partial^2f(\phi_{\mu\nu})}{\partial\phi_{\mu\nu}^2}\right\rvert_{\phi_{\mu\nu}=0},
\end{equation}
where $f$ is the free energy density and $\phi_{\mu\nu}$ is an external flux through the $(\mu,\nu)$ planes.
In a massive (confining) phase, $h$ goes to zero exponentially with the system size.
In a massless (Coulomb) phase, it can be non-zero in the thermodynamic limit.
Thus, it can be used as an order parameter in the $4d$ $U(1)$ theory, as shown
in \cite{Vetto}. 
While $h$ can be expressed as a simple expectation value with the Wilson action,
with the topological action it cannot. As in \cite{hel_mod}, we measured the
distribution of allowed twist angles $\phi_{\mu\nu}$, fitting the curvature of the corresponding
effective potential to obtain $h$.

\begin{figure}
\centerline{
\includegraphics[width=0.50\linewidth]{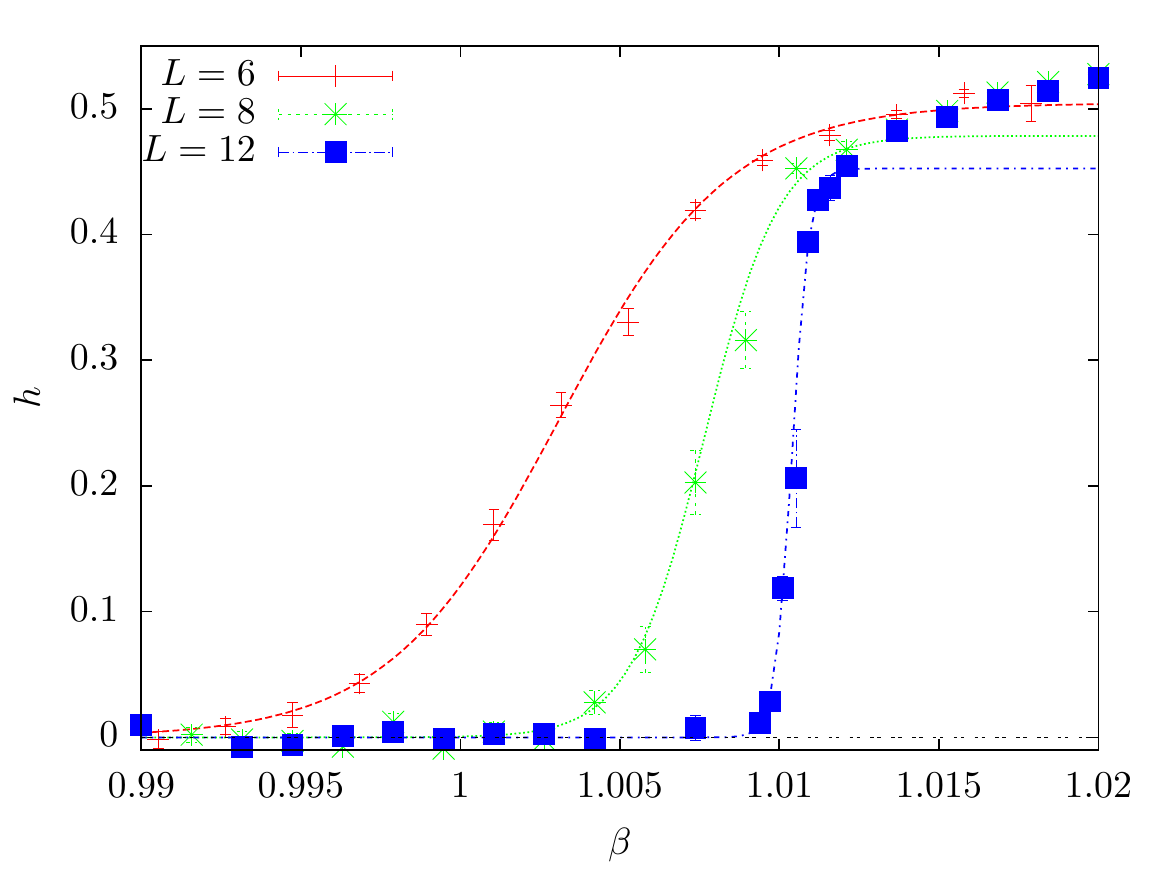} 
\includegraphics[width=0.50\linewidth]{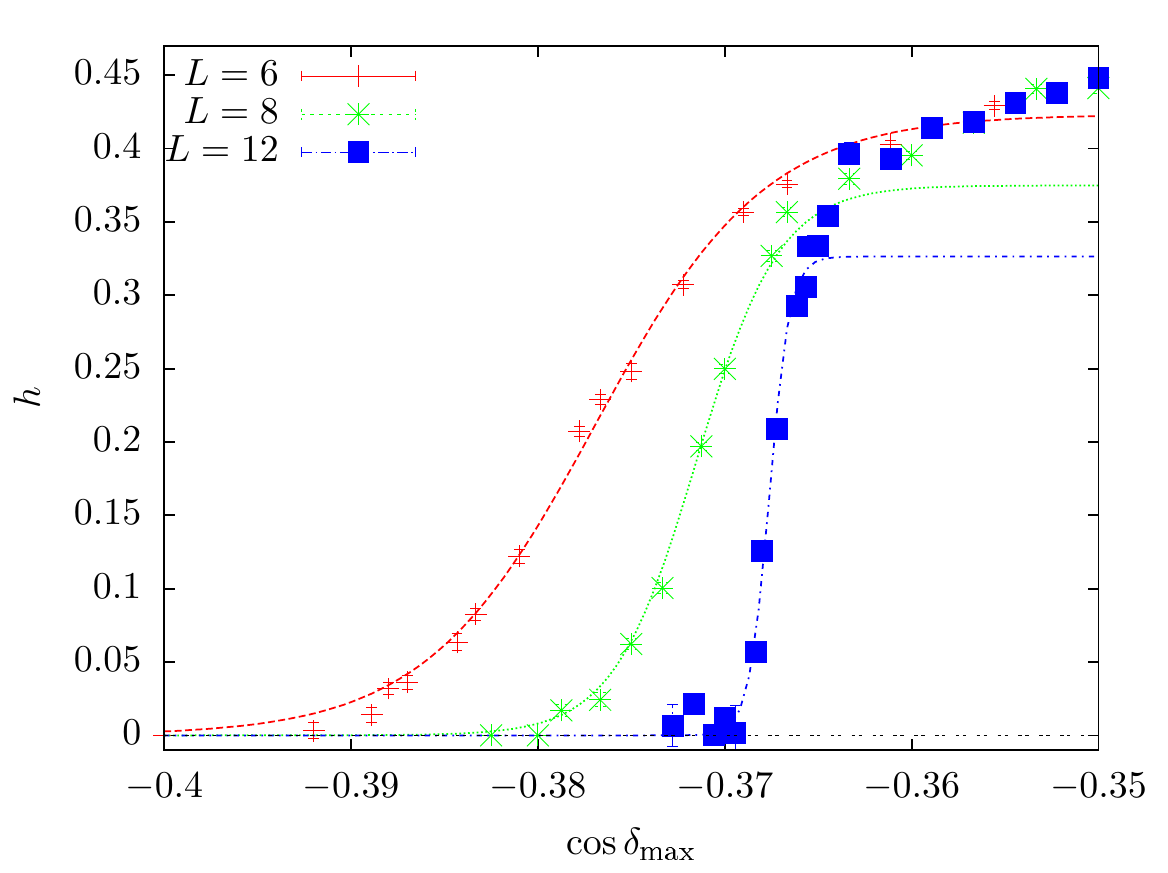}
}
\centerline{
\includegraphics[width=0.50\linewidth]{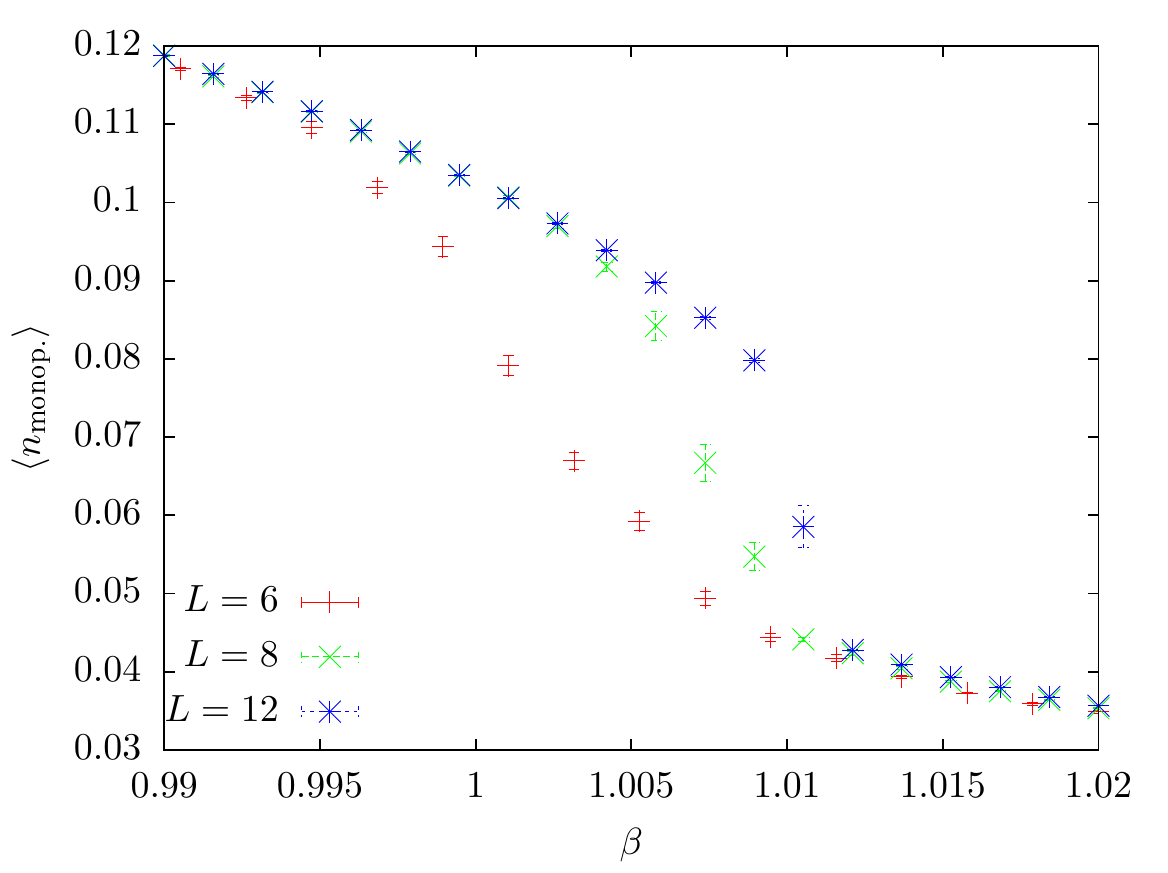} 
\includegraphics[width=0.50\linewidth]{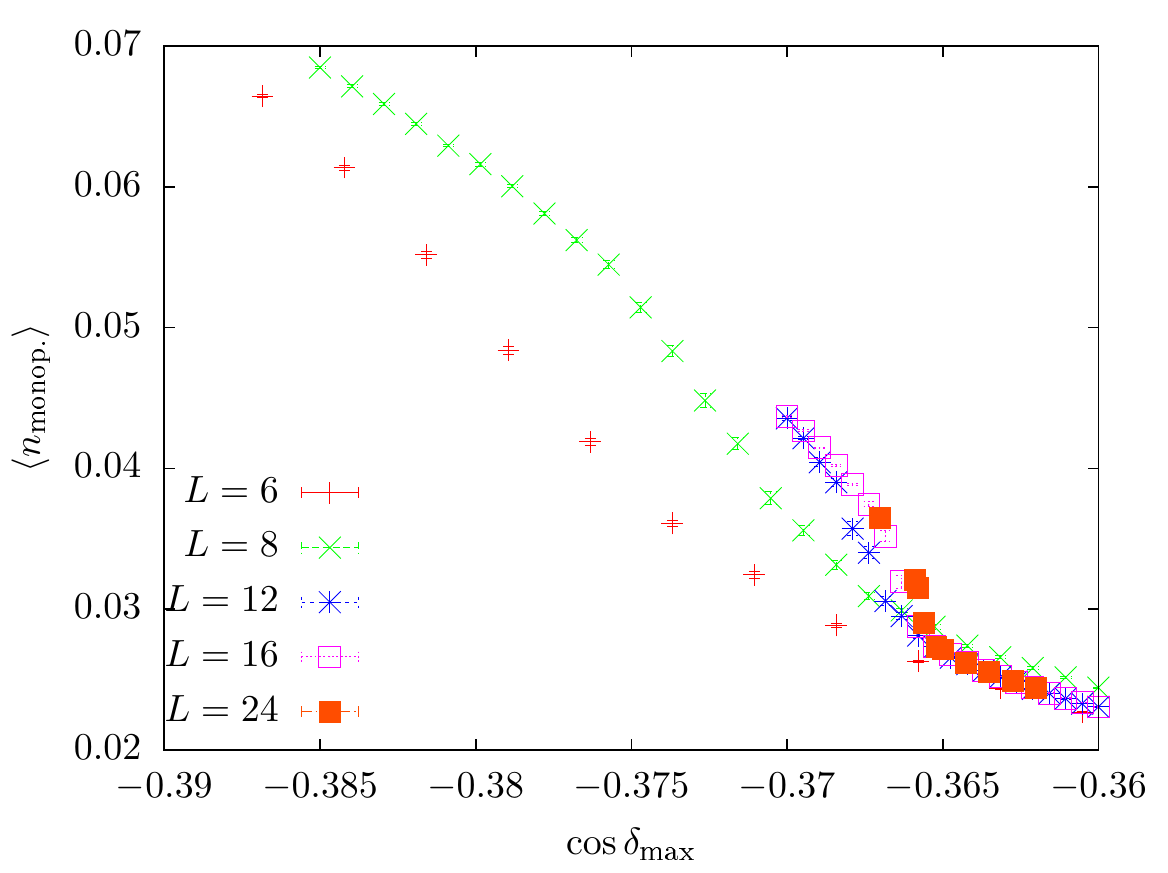}
}
\caption{Helicity modulus ({\em top row}) and monopole density ({\em 
bottom row}) for the Wilson action ({\em left column}) and for the
topological action ({\em right column}), across the phase transition.
For both actions, the transition is first-order: the helicity modulus 
jumps at the transition and is well described by a finite-size first-order
ansatz~\cite{Borgs}; the monopole density also jumps, but the jump is smaller for
the topological action.
}
\end{figure}

In Fig.~2 (top), we show the measured values of the helicity modulus,
as a function of $\beta$ for the Wilson action (left), and as a function of
$\cos \delta$ for the topological action (right). The lines are fits assuming
a first-order transition, where the free energy varies linearly with the
control parameter (but with different coefficients in the two phases)~\cite{Borgs}.
The similarity between the two figures leaves no doubt that a first-order
transition (confining $\leftrightarrow$ Coulomb) takes place with the
topological action, at a cutoff angle $\cos\delta_c \approx -0.368$ quite
different from $\delta=\pi/3$.
A measurement of Creutz ratios in both phases is consistent with an area
law and a Coulomb law, respectively.

Note that the helicity modulus in the Coulomb phase has almost the same value
at the transition for both actions. On the other hand, the fitted latent heat
is about half as large with the topological action.
Also, the monopole density changes discontinuously at the phase transition,
for both actions, as illustrated Fig.~2 (bottom), but the topological action
yields a smaller density and associated jump  (right).

In addition to being an order parameter, the helicity modulus leads also
to a simple definition of the renormalized charge~\cite{Vetto}. The free 
energy $f(\phi_{\mu\nu})$ is well described by the classical ansatz
\begin{equation}
f(\phi_{\mu\nu}) = -\log\sum_ke^{-\frac{\beta_R}{2}\left(\phi_{\mu\nu}-2\pi k\right)^2}
\end{equation}
from which $\beta_R = 1/e_R^2$ can be extracted, for either type of action.

\section{Monopole density in the Coulomb phase}

In the confining phase, monopoles condense. In the Coulomb phase, monopoles
are exponentially suppressed: the density of monopole current behaves
as $\exp(-\beta_R M)$, where $M$ can be considered as the monopole mass.
Fig.~3 (left) shows this exponential suppression, for both actions.
With the topological action, monopoles are much heavier, i.e. more suppressed.
This is evidence that the topological action is {\em improved}: for a given
value of the renormalized coupling, lattice artifacts are reduced considerably.

Actually, as explained in Sec.~2, with the topological action monopoles
cannot exist if the cutoff $\delta$ on the plaquette angle is $\pi/3$ or less.
Thus, the density of monopole current vanishes, and is singular at 
$\delta = \pi/3$. We tried to monitor this singular behavior. As $\delta$
approaches $\pi/3$, the density of monopole current deviates from the 
exponential behavior of Fig.~3 (left) and seems to obey a power law,
being proportional to $(\cos(\pi/3) - \cos(\delta))^\gamma$.
See Fig.~3 (right). We measure a large value $\sim 12$ for 
the ``critical exponent'' $\gamma$, which can be explained heuristically.

Consider first a $2d$ $XY$ model, with topological action
$\exp(-S_{\rm top}) = \prod_{\langle ij \rangle} \Theta(\delta - |\theta_i - \theta_j|)$. Vortices in this model come from the spin orientation $\theta_i$
winding by $2\pi$ as one goes around an elementary plaquette. Thus,
vortices cannot exist if $\delta < \pi/2$. When $\delta$ just exceeds $\pi/2$,
each angle $|\theta_i - \theta_j|$ around a vortex must be near $\pi/2$,
within an angle $(\delta - \pi/2)$ of it. The phase space for this is
$\propto (\delta - \pi/2)$. Since there are 4 links around a vortex,
related by one constraint (the sum of the four $(\theta_i - \theta_j)$
is $2\pi$), the phase space for a vortex is $\propto (\delta - \pi/2)^3$.
Now, on a periodic lattice a vortex is always accompanied by an anti-vortex.
Thus, the probability of finding a vortex anti-vortex pair should be
$\propto (\delta - \pi/2)^6$. This is remarkably consistent with the
measurements shown Fig.~4 (right). 
We can then apply a similar argument to the $U(1)$  monopole density. 
The smallest monopole current loop is dual to 4 cubes. 
Each cube contains 6 plaquettes, which together obey a Bianchi constraint
so that the total outgoing flux through the 6 plaquettes is $2\pi$.
Thus, the probability of finding such a monopole loop is
$\propto (\delta - \pi/3)^{4\times 5}$. Our data, shown Fig.~4 (left), is
more consistent with an exponent $\sim 12$, indicating that our heuristic
argument is too crude.

\begin{figure}
\centerline{
\includegraphics[width=0.50\linewidth]{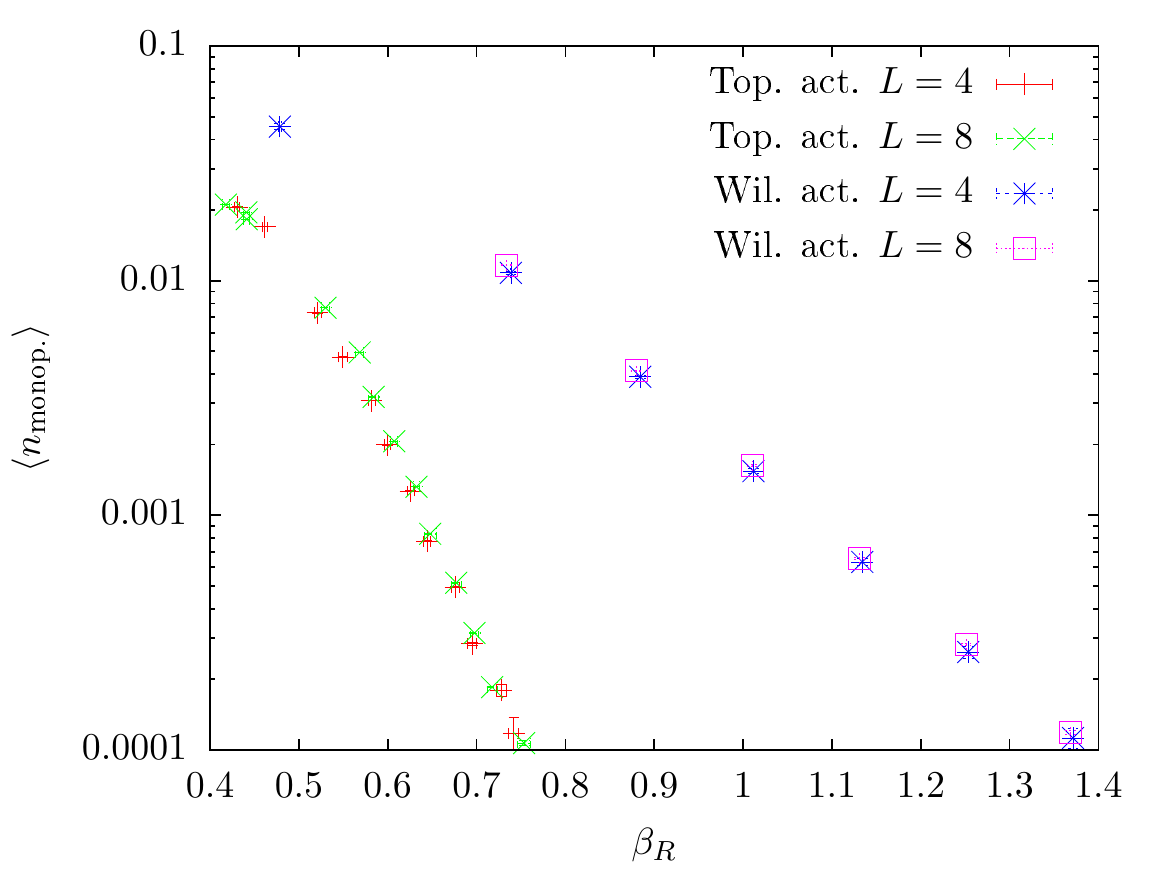} 
\includegraphics[width=0.50\linewidth]{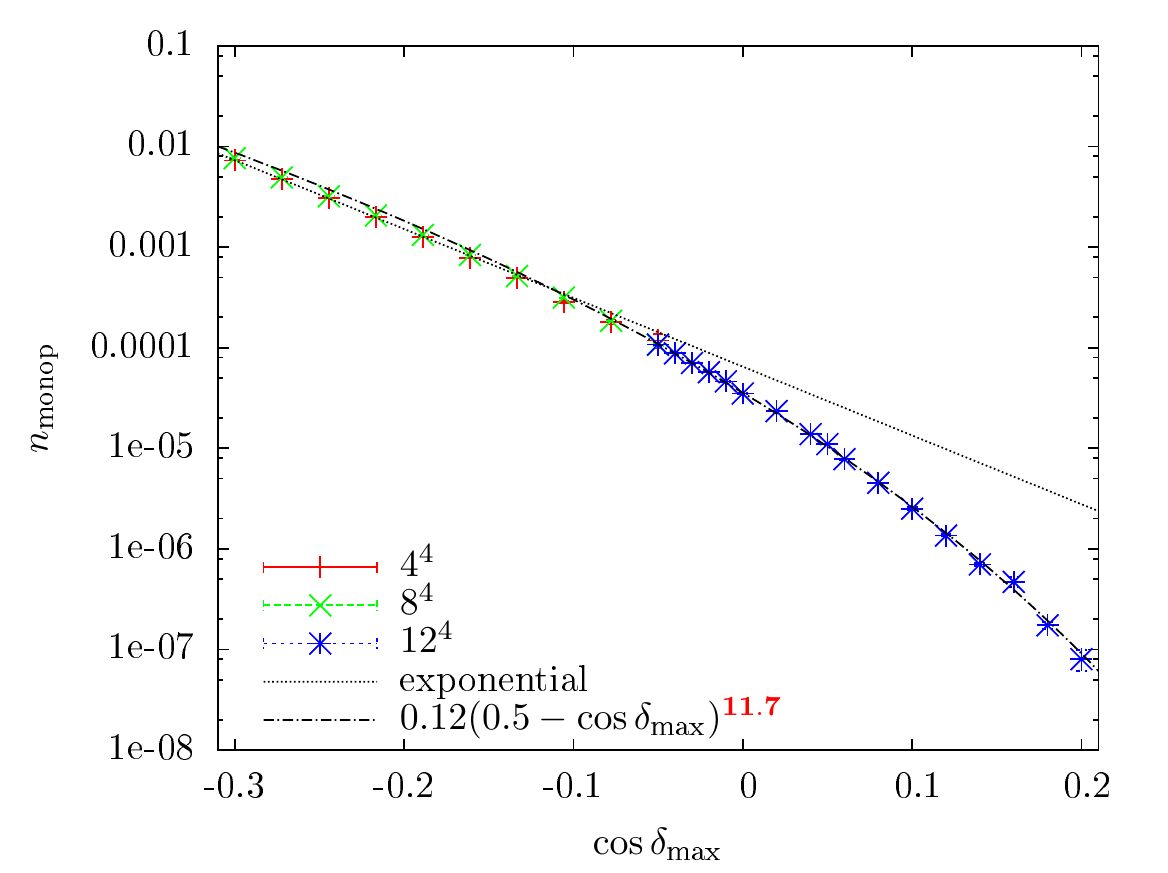}
}
\caption{Monopole density in the Coulomb phase. {\em Left}: The monopole
density decreases exponentially as a function of the inverse renormalized 
coupling $\beta_R = 1/e_R^2$, both for the Wilson action and for the 
topological action, with a steeper decrease for the latter. {\em Right}:
using the topological action, the monopole density vanishes when the
plaquette angles cannot exceed $\pi/3$. 
As one approaches this critical value,
the monopole density vanishes as a power law of the distance to the
critical constraint angle, with a critical exponent $\sim 12$.
}
\end{figure}

\begin{figure}
\centerline{
\includegraphics[width=0.50\linewidth]{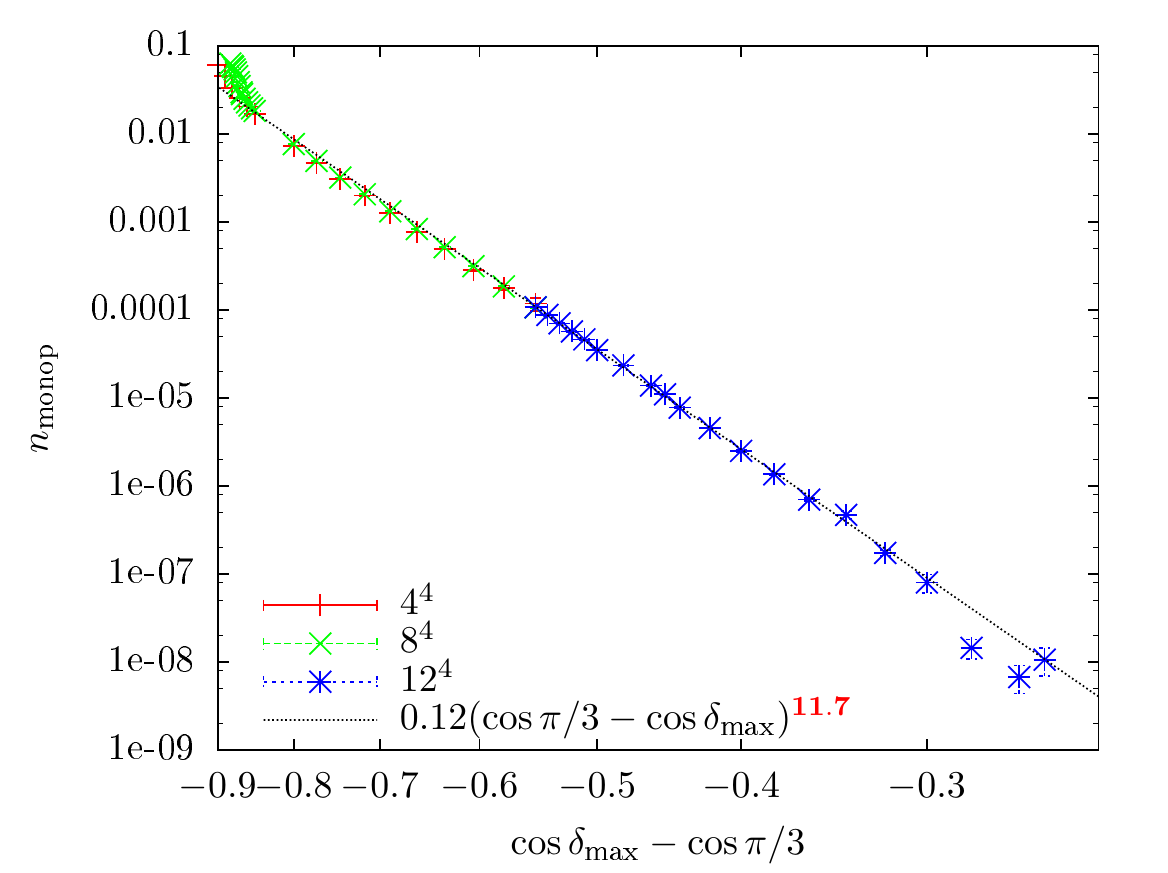} 
\includegraphics[width=0.50\linewidth]{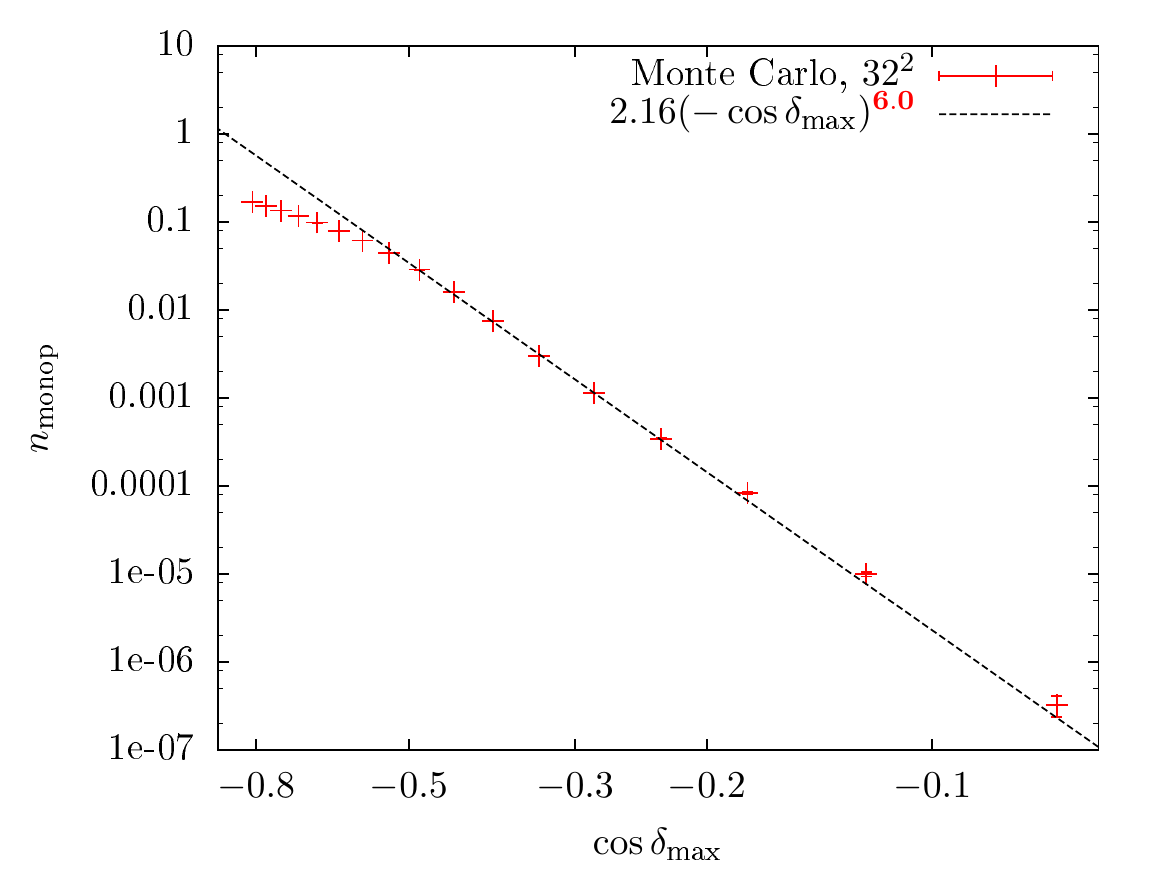}
}
\caption{{\em Left}: monopole density in the Coulomb phase, for the 
topological action, as a function of the distance to the critical constraint:
a power law is clearly visible on this log-log plot.
{\em Right}: similar figure showing the density of vortices in the massless
phase of the $2d$ XY model, using a topological action. The vortex density
vanishes as a power of the distance to the critical constraint, which limits
the difference among neighboring spin angles to $\pi/2$. The critical exponent
6 is consistent with a heuristic argument (see text).
}
\end{figure}

\section{Free energy for free}

With a topological action, the partition function $Z(\delta)$ simply counts the number of configurations
which satisfy the constraint. Therefore, $Z(\delta_2)/Z(\delta_1)$, with $\delta_2 < \delta_1$,
is the proportion of configurations satisfying the constraint $\delta_1$ which also satisfy the
tighter constraint $\delta_2$. These configurations can be counted as the Monte Carlo
sampling of $Z(\delta_1)$ proceeds, giving the variation of the free energy $f(\delta)$
``for free''. 

We have applied this simple strategy near the phase transition, in the $2d$ $XY$ model
and in the $4d$ $U(1)$ gauge theory. The phase transition is of infinite order (BKT) in
the former case, of first-order in the latter. The results for $\partial f(\delta)/\partial\delta$,
shown Fig.~5, make this difference clear: for the first-order transition (left), the first
derivative of the free energy develops a discontinuity as the system size increases, 
while it remains smooth for the infinite-order transition.

\begin{figure}
\centerline{
\includegraphics[width=0.50\linewidth]{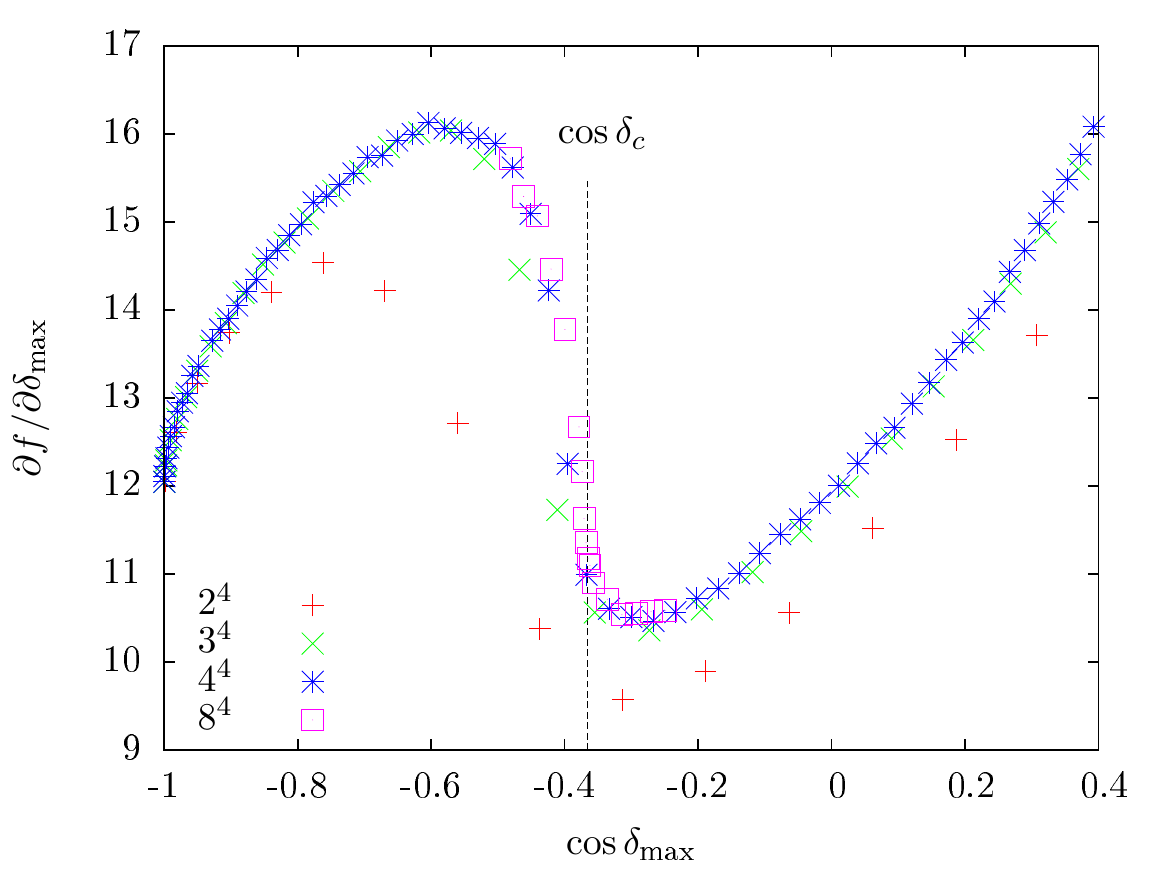} 
\includegraphics[width=0.50\linewidth]{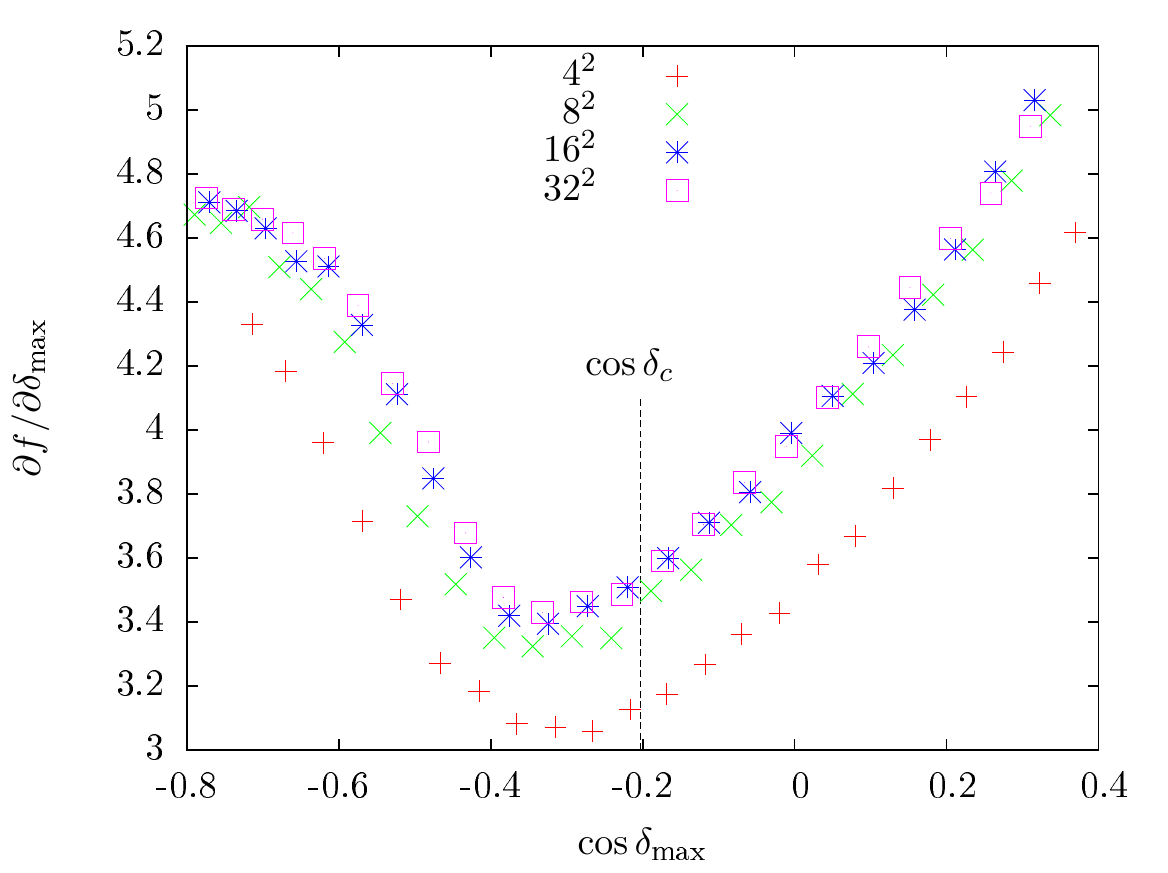}
}
\caption{Derivative of the free energy density with respect to the constraint 
angle,
across the confinement-Coulomb phase transition marked by the vertical line, for the $4d$ $U(1)$ gauge
theory ({\em left}) and for the $2d$ $XY$ model ({\em right}).
As the system size is increased, a discontinuity is clearly visible in 
the former case, signaling a first-order transition. In contrast, the
derivative of the free energy remains smooth in the $XY$ model, as
expected from a transition of infinite order.
With a topological action, the free energy is obtained by simply counting
the number of configurations which survive a tighter constraint (see text).
}
\end{figure}

\section{Conclusion}

A topological action takes only two values: $0$ and $+\infty$, where the latter removes any configuration
which violates a local smoothness constraint. We have studied the properties of the topological action
$S(\delta) = -\sum_P \log \Theta(\delta - |\theta_P|)$ in the $4d$ $U(1)$ gauge theory, as a function of
the constraint angle $\delta$. As $\delta\to 0$, all plaquettes $P$ are forced to approach the identity.
Yet the approach to the continuum action $\int d^4x F_{\mu\nu}^2$ is not clear a priori.
Nevertheless, we have shown that the phase diagram of this lattice theory is very similar to that obtained
with the Wilson action, with a confining, disordered phase at large $\delta$, separated from a
Coulomb, ordered phase at small $\delta$ by a first-order transition. 
This transition is associated with a jump in the density of magnetic monopoles, which condense
in the confining phase and are exponentially suppressed in the Coulomb phase. 
With the topological action, the monopole density is smaller in the Coulomb phase, which
reflects the improvement of the action.
The helicity modulus serves
both as an order parameter for the phase transition, and as a renormalized coupling in the Coulomb
phase.

A feature specific to the topological action is the total suppression of monopoles, or of topological
defects in general, for a particular value of the constraint (here, $\delta_c=\pi/3$). This threshold
is analogous to L\"uscher's ``admissibility condition'', which guarantees that topological sectors
are distinct, even on the lattice~\cite{Luscher}. Here, the monopole density is singular at $\delta_c$: we observe
that it decreases as a high power of $(\delta - \delta_c)$ when $\delta\to\delta_c^+$, then is
identically zero for $\delta < \delta_c$. Other effects of this phase transition, if any, remain
to be elucidated.

Finally, a topological action provides a simple means to measure the free energy, as the fraction
of configurations in a given ensemble which continue to satisfy the constraint as this constraint
is tightened. This might be useful in some attempts to tackle the sign problem, where the
free energy must be convoluted with a complex phase factor~\cite{Langfeld}.

\end{document}